\documentclass{aa}
\input{epsf.tex}
\newcommand{\ltapprox}{\raisebox{-0.5ex}{$\,\stackrel{<}{\scriptstyle\sim}\,$}}

\begin{document}
 
   \thesaurus{03(11.05.1; 11.19.2; 11.12.2)}
%
\title{Near--infrared Luminosity Function in the Coma cluster
\thanks{Based on observations collected with the {\it T\'elescope
Bernard Lyot}, at the Pic du Midi Observatory, operated by INSU (CNRS).}
}
 
\author{ S. Andreon\inst{1}, R. Pell\'o\inst{2}}

   \institute{
Osservatorio Astronomico di Capodimonte, via Moiariello 16,
80131 Napoli, Italy (email: andreon@na.astro.it)
\and
Observatoire Midi-Pyr\'en\'ees, LAT, UMR 5572,
14 Avenue E. Belin, F-31400 Toulouse, France 
(email: roser@obs-mip.fr)
}

   \date{Accepted}
   \titlerunning{$H$ band Coma luminosity function} 
   \maketitle 
 
   \begin{abstract} 
We present the near--infrared $H$ band luminosity function (hereafter LF)
of the Coma cluster of galaxies. It is the deepest ever computed in the
near--infrared, for any
type of environment, extending over 7 magnitudes, down to $\sim M_H^*+6$.  
The LF was computed on a near--infrared selected sample of galaxies which
photometry, complete down to the typical dwarf luminosity, is presented in
a companion paper. The Coma LF can be described by a Schechter function
with intermediate slope ($\alpha\sim-1.3$), plus a dip at $M_H\sim-22$ mag.
 The shape of the Coma LF
in $H$ band is quite similar to the one found in the $B$ band and, with
less confidence, to the $R$ band LF as well.  The similarity of the LF in
the optical and $H$ bands implies that in the central region of Coma there
is no new population of galaxies which is too faint to be
observed in the optical band (because dust enshrouded, for instance), down
to the magnitudes of dwarfs. The exponential cut of the LF at the bright
end is in good agreement with the one derived from shallower near--infrared
samples of galaxies, both in clusters and in the field. This fact is
suggestive of a similarity of the tip of the mass function of galaxies,
irrespective of the environment where they are found. The dip at
$M_H\sim-22$ mag is instead unique among all the so far measured
near--infrared LF, although several published
observations are not deep enough or spanning a suitable wide field to
distinctly detect this feature. The faint end of the LF, reaching
$M_H\sim-19$ mag (roughly $M_B\sim -15$), is steep, but less than previously
suggested from shallower near--infrared observations of an adjacent region
in the Coma cluster. The differences between our measured LF and that measured
previously in other regions suggests a dependency on environment of the 
faint end of the mass function (below $M^*+2.5$).

\end{abstract}
 
\section{Introduction}

In the optical band, the cluster luminosity function (hereafter LF) has
three regimes:  a bright end ($M_r^*\sim-22$ mag, $H_0=50$ km s$^{-1}$
Mpc$^{-1}$), a flat slope ($\alpha\sim-1.0$) down to luminosity of dwarf
galaxies (L\'opez--Cruz et al. 1997; Garilli, Maccagni \& Andreon 1999),
and then a steep increase (Impey, Bothun \& Malin 1988; Ferguson 1989;  
Thompson \& Gregory 1993;  Secker \& Harris 1996;  Secker, Harris \&
Plummer 1997). Often a dip is found in the otherwise flat part of the LF
(see, e.g., Godwin \& Peach 1977; Bucknell, Godwin \& Peach 1979). A few
well studied clusters (Smith, Driver \& Phillipps 1997), as well as number
of LFs published a long time ago (Schechter 1976)  display an intermediate
slope ($\alpha\sim-1.3$) instead of a flat LF.

The LF represents the zero--order statistics of galaxy samples 
and gives the relative number of galaxies as a
function of the magnitude. Almost every quantity is, therefore,
``weighted" by the LF, including obvious quantities, such as the galaxy
color distribution, and also less obvious ones, such as correlations 
involving the luminosity (see,
for example, the discussion on the impact of magnitude limits in the
size--luminosity relation by Simard et al. 1999). When the sample is not
complete in volume a further ``weight" should be added: the selection
function.  Thus, an accurate knowledge of the LF is important when
comparing galaxies of different luminosities at different redshifts.

From a physical point of view, the optical LF is the convolution of the
number of galaxies of a given mass with their M/L distribution. Then, any
measure of the optical LF traces a complex mix of galaxy mass and M/L
distributions, so that evolution in luminosity or mass could not be easily
disentangled from the measurement of the optical LF. A better estimate of the
galaxy mass than the optical luminosity will certainly help to separate the
two dependencies. Such a measure has a particular relevance in the
determination of the density of the Universe: a possible way to proceed is to
compute the cluster mass per unit luminosity times the Universe luminosity
density. As stressed by Calberg et al. (1996), this calculation assumes that
cluster galaxies have the same LF as field galaxies.  Observations suggest
instead that galaxies change their optical luminosity during their infall in
the cluster (see, for example, Bothun \& Dressler 1986;  Andreon 1996, and
most of the papers by the CNOC collaboration, such as Balogh et al. (1998)
and references therein), although the amplitude and the sign of the
luminosity variation is not yet settled.  Of course, it would be preferable
to measure the M/L of clusters using a luminosity indicator weakly 
affected by possible bursts or halt of star formation induced by interactions
with the hostile cluster environment.

The near--infrared luminosity has several advantages with respect to optical
luminosities. It is tightly correlated to the
galaxy mass (at least for spirals, Gavazzi, Pierini \& Boselli 1996)  and,
with respect to the optical luminosity, it is less affected by short and
recent star formation events (Bruzual \& Charlot 1993), possibly induced by
interactions, and by dust absorption. Therefore, the near--infrared LF
traces more directly the mass function and gives a Universe density less
affected by possible systematic errors due to a differential star formation
history between galaxies in clusters and in the field.

There are several additional advantages in observing galaxies in the
near--infrared: K corrections are relatively small and well known, thus
allowing to observe and to compare galaxies at different redshifts, up to
high redshift values. In particular, K corrections are almost independent
from the spectral type of galaxies, in such a way that statistics on a
population of galaxies are less affected by changes of the morphological
composition induced by differential corrections from type to type.
Furthermore, galaxies that undergo a starburst are not selected
preferentially, as instead happens in the optical, and therefore a sample
selection in the near--infrared is less biased by episodic events of star
formation.

It is therefore important to measure the near--infrared LF of clusters of
galaxies over a magnitude range as wide as possible, in particular to
characterize the properties of galaxies in the local Universe. So far, the
near--infrared LF have been measured for a few clusters, but to bright
limiting magnitudes (Barger et al. 1996, Trentham \& Mobasher 1998,
De Propris et al. 1999), and on
a portion of the Coma cluster (De Propris et al. 1998), down to relatively
faint magnitudes. According to De Propris et al. (1998), the Coma LF shows a flat
slope, and a step increase ($\alpha\sim-1.7$) at faint magnitudes ($H=16$
mag). However, this is presently the only LF determination attaining
intermediate magnitudes, and such a survey could be improved in several
respects. It is important to extend the study to other regions, and to reach
deeper magnitudes. This is the aim of the present paper.

We present the near--infrared LF of the Coma cluster, based on independent
observations, fully documented in a companion paper that also presents the
photometric catalog.  With respect to De Propris et al. (1998), this study
has been performed on a different portion of the Coma cluster, slightly
overlapping with their one, over an area which is $\sim$ 40 \% smaller, but
it attains one magnitude deeper.  All along this paper, we adopt $H_{0}=50$
km s$^{-1}$ Mpc$^{-1}$ and q$_{0}=$ 0.1.


\section{Data analysis} 

Coma galaxy counts and LF have been computed from the photometry presented
in Andreon, Pell\'o, Davoust et al. (1999), to which we defer for details
(hereafter Paper I). In summary, a $\sim 20 \times 24$ arcmin region
of the Coma cluster, located $\sim 15$ arcmin from
the centre, have been imaged with the Moicam camera at the 2.0m Bernard 
Lyot telescope at Pic du Midi. Images were taken in the $H$ band under
moderate to good seeing conditions (i.e. $1<FWHM<1.5$ arcsec), with average
exposure time of $\sim300$ sec. About 300 objects have been detected and
classified by Sextractor version 2 (Bertin \& Arnouts 1996) in the best
exposed part of our mosaic ($\sim380$ arcmin$^2$).

Different magnitudes are presented in Paper I.  We adopt here the Kron
magnitudes (see Kron 1980 for the exact definition, and Bertin \& Arnouts
1996 for the software implementation). They are defined as the flux
measured in a region which area is adapted to each galaxy.  Unfortunatly,
they depend sensibly on the determination of the object size, in particular
for faint objects, and therefore, for faint objects we prefer aperture
magnitudes. More precisely, we adopt, as a measure of the magnitude for a
galaxy, the magnitude computed within 2.5 Kron radii for galaxies brighter
than $H=14$ mag, and aperture magnitudes (within 10 arcsec aperture) for
fainter galaxies. The two quantities are identical, within the errors, for
galaxies in a large magnitude range including $H\sim14$ mag (Paper I). The
catalog is complete, in the 10 arcsec aperture magnitude, down to
$H=17.1$--$17.2$ mag\footnote{All magnitudes are refered to the Vega system}. 
To be safe and for easy computation, we cut the
catalog at $H=17.0$ mag. Given the galaxy catalog and the knowledge of
the surveyed area, galaxy counts in the Coma direction are computed
straightforwardly. They are presented in Figure 1 (open dots), as derived
for objects identified as galaxies (see Paper I for details).
 
The Coma cluster LF is computed as the statistical excess of galaxies in the
Coma cluster direction with respect to other directions. In order to
estimate the fore and background contribution of the field, we use when
possible observed values measured by different authors, as well as a simple
standard model for number counts, based on pure luminosity evolution for
galaxies (see Rocca-Volmerange \& Guiderdoni, 1990, Pozzetti et al. 1996 and
Pozzetti et al. 1998) and computed through the Bruzual \& Charlot
evolutionary code (1993, updated as GISSEL98). The parameters of the counts
model have been set up in order to roughly reproduce the observed number
counts to B = 28 mag (Williams et al. 1996), and normalized to the observed
counts at $H=17.0$ mag. 
This model is only used in order to derive the mean redshift of the
dominant population at a given magnitude, when comparing with other 
LF estimates, computed with other filters.

Field counts have been measured on the $H$ band images of the Hubble Deep
Field South 1 \& 2 (hereafter HDFS1+S2), presented in da Costa et al.
(1999). These images were taken at the NTT and they are much deeper (several
magnitudes) than our Coma images, but extending to a smaller region and
exposed in
an non uniform way. We have used their uniformly exposed part, a central
region $1000 \times 1000$ pixel wide (i.e.  $\sim 23.7$ arcmin$^2$ large,
thus more than 10 times smaller than the Coma area studied in this paper).
We have detected and classified objects in this HDFS1+S2 area by means of
Sextractor (Bertin \& Arnouts 1996), using the same parameters as in 
Paper I. Figure 1 presents the resulting counts (closed triangles).  
At $H\sim16$ mag, field counts have large errors because the HDFS1+S2 is not
tailored for measuring galaxy counts at such bright magnitudes, but for
going deep on a small region. This fact prompt us to look for a $H$ band
survey more adapted to our aims, i.e. shallower and wider. Since it does not
exist, we get a different estimate of the $H$ band galaxy counts using $K$
band galaxy counts and assuming a mean $H-K$ color for galaxies in the
relevant magnitude range.  The observed color of $H\sim17$ mag galaxies is
$H-K\sim 0.6$ mag (Stanford, Eisenhardt \& Dickinson 1995). This value is also
in fairly good agreement with the mean $H-K$ expected from the counts model
($H-K\sim 0.55$ mag to $H\sim19$ mag, where the population is dominated by galaxies
with $0.1 \ltapprox z \ltapprox 0.4$ at $H\sim17$ mag, and with $0.2 \ltapprox z
\ltapprox 0.6$ at $H\sim18 - 19$ mag). We apply the mean $H-K$ value to the
Bershady, Lowenthal \& Koo (1998) compilation of $K$ band surveys. These
counts are presented in Figure 1 as a strip with center given by the average
counts presented in their paper. There are also presented in Figure 1 the
expected number counts derived from our model (solid line histogram). In
spite of unavoidable differences between the types of magnitude used by the
different authors, and also the approximations involved in the conversions
between photometric systems, the agreement between the galaxy counts in the
HDFS1+S2 direction and the $H$ counts estimated from $K$ counts is very
good. They are also in good agreement with the counts derived form our simple
model. We adopt these counts as average background counts in the Coma
direction.
 
An expected and important source of error in the LF determination is the
background variance from field to field, in addition to Poissonian fluctuations: 
if the background variance is high,
then the background in the Coma direction could be significatively different
from the average computed above. Among the $K$ band shallow surveys, two of
them are adapted to roughly compute the order of magnitude of this variance.
Gardner et al. (1993) presented galaxy counts for the HMDS (Hawaii Medium
Deep Survey) extending over an area which is only half of that sampled for
Coma, and also for the HMWS (Hawaii Medium Wide Survey), extending over a
much wider area than the present one. The counts in the two surveys show a
$\pm15$ \% scatter, and we adopt this value as a typical fluctuation for the
background (to be added quadratically to Poissonian fluctuations).  The
amplitude of the strip in Figure 1 shows this scatter. This background
variance seems plausible for two reasons: first, counts in the HDFS1+S2,
which extend on an area $\sim10$ times smaller than our one, and $\sim3$
times smaller than the HDMS survey, are well within the strip, showing that
background fluctuations are unlikely to be larger than our derived variance.
Secondly, the expected field to field fluctuations are $\sim11$ \%,
according to the formulas (and the hypothesis) in Huang et al. (1997).

Figure 1 shows that at all magnitudes considered here, the Coma cluster
counts have small errors and stand out with respect to the field counts,
down to $H=17$ mag. Therefore, errors on the
Coma LF will be small and only slightly affected by the background
subtraction. In order to judge on the progress achieved in this paper with
respect to previous investigations, the reader can compare our
magnitude--counts diagram with the analogous one in Mobasher \& Trentham
(1998) for a much smaller (and denser) region of Coma. These authors took an
observational strategy quite different from our one: given the available
telescope time, they went as deep as possible on a very small area, which
resulted in a large field to field background variance.
 
\section{Results}

\subsection{The shape of the LF}

Given the counts in the Coma direction and in the field, and their errors,
the computation of the Coma cluster LF is an algebrical exercise. We stress
that our main sources of error on the Coma LF are Poissonian fluctuations of
total counts over the Coma region and the background field to field
variance. Therefore, the error on the background is not derived from the
HDFS1+S2 errorbars (which are not relevant for the determination of the LF
errors). 

The near--infrared Coma LF is presented in Figure 2. It is characterized by
a bright end (at $M_H\sim-25$ mag), a part increasing gently down to
$M_H\sim-18.5$ mag, and an ``outlier" point at $M_H=-22.2$ mag, which
produce, if real, a dip in the Coma LF.  The LF displayed in Figure 2 is the
deepest ever measured in any near--infrared band for any type of
environment for a near--infrared selected sample.

In the optical, the Coma cluster exhibits a similar shape (Godwin \& Peach
1977; Secker \& Harris 1996).  Using Godwin, Metcalfe \& Peach (1983) data,
we computed the $b$ (a photographic $J$-like blue filter) Coma LF in almost
exactly the same area surveyed in the $H$ band. For simplicity, we have
considered a rectangular area enclosing our $H$ band region, without taking
into account the complex geometry of our region in details. Then, because
the $H$ and $b$ band magnitudes are available for all galaxies in this area,
we have computed a mean $b-H$ pseudo-color\footnote{As long as the two
magnitudes are not computed within the same aperture, the difference is not
actually a color.}. In order to compare the two LFs, we have shifted the $b$
band LF by the mean $b-H$ pseudo-color: $b-H = 3.5$ mag. The expected values
according to Bruzual \& Charlot models are $b-H = 4.0$ for elliptical
galaxies and up to $b-H = 2.8$ for blue constant star-forming systems; thus
an averaged population of $2/3$ of ellipticals versus $1/3$ of blue systems
gives roughly the right mean value as expected. No normalization in $\Delta
n$ has been applied. The result of this exercise is shown in Figure 2 as a
dashed-line histogram. The two LFs are remarkably similar:  there is a close
agreement between the cut at $M_H=-25$ mag, the dip location ($M_H=-22.2$ mag),
the dip amplitude and the increase observed at fainter magnitudes (closed
dots) and expected from the $b$ LF (dashed histogram). We have performed
the same exercise with the $R$ band photometry by Secker \& Harris (1996),
who studied an adjacent region of the Coma cluster. The expected values
according to Bruzual \& Charlot models are $R-H=2.5$ for early--type
galaxies and $R-H=2.0$ for blue star-forming systems, giving an averaged
value of $R-H=2.3$ for the same weighted population taken above ($2/3$ of
ellipticals versus $1/3$ of blue systems). This time we adopt the predicted
value because the $R$ catalog is not published. The two surveyed regions
are different, thus we normalize their LF to our $H$ LF.  We obtain similar
results, with the difference that the importance of the dip is smaller in
$R$ than in $H$ (see Figure 2).

Thus, the shape and the amplitude of the Coma LF seems not to be strongly
depend on the wavelength when we compare the results in $b$ and in $H$
bands, and also, with less confidence, in the $R$ band. The strong similarity of
the optical and near--infrared LF implies that in the near--infrared there
is no new population of galaxies which disappears in the optical band
(because dust obscured, for example), down to the magnitude of dwarfs.
Furthermore, if the $H$ band LF traces the galaxy mass function in this
cluster, the same holds true for the blue LF.  This result has been obtained
in a particular region of Coma, wich is a cluster rich in elliptical and
lenticular galaxies. Before any generalization, this result should be
checked in other regions of the cluster, and also in other environment
conditions (cluster outskirts, clusters rich in spiral galaxies, groups, ...).

\subsection{The dip at $M_H \sim -22 $}

Let us consider in more details the dip point. The question is: Is it really
an outlier? The statistical significance of the possible outlier point must
be evaluated from the galaxy counts, since they are the original source of
fluctuations. The $M_H=-22.2$ mag bin in Figure 2 corresponds to the
$H=13.5$ mag bin in Figure 1. First of all, we exclude the possibility that
we have missed some galaxies of this magnitude, because we are complete 2.5
mag fainter, and because a typical galaxy of $H=13.5$ mag have a central
brightness of 100 times the sky noise. Secondly, there is no relation
between the location of the dip and the discontinuity of the magnitude
system adopted (Kron magnitudes for bright galaxies and aperture magnitudes
for faint ones): galaxy counts does not change because the separation is set
to $H=14$ mag.  In fact, Kron and aperture magnitudes have almost the same
value down to the last magnitude bin (see figure 10 in Paper 1).  There are
two galaxies in the $H=13.5$ mag bin, whereas $\sim13$ are needed to make
the counts smooth. Therefore, this point
is more than 3 $\sigma$ away from the average of adjacents bins. To be
precise, according to Poissonian statistics
we can reject at more than 99.95 \% confidence level the hypothesis
that the observed number of galaxies is drawn from a parent distribution
which counts $\sim13$ galaxies in that bin. Therefore, the dip is a real
feature of the Coma near--infrared LF in this region.


The dip in the Coma LF had firstly been noticed in the optical band (for
example, Godwin et al. 1983) and it had been interpreted in two different
ways. Biviano et al. (1995) suggested that galaxies brighter than the dip
were subjected to a recent episode of star formation induced by
the hostile Coma environment, which have made them brighter. Andreon (1998)
has shown that the LF of the different morphological types of galaxies are
equal in Coma and in much poorer environments, and that the dip is simply
the combined result of the Coma cluster morphological composition together 
with the shape of the type--dependent LFs. If the induced star-formation 
interpretation by Biviano et al. (1995)
was correct, the dip should be absent or at least highly
attenuated in $H$, because the near--infrared luminosity traces the galaxy
mass and it is less affected by the short timescale starbursts that makes a
few galaxies brighter than the magnitude of the dip. Instead the dip is
observed in the $H$ band.

A few other near--infrared LF of clusters are (poorly) known. None of the
five clusters studied by Mobasher \& Trentham (1998) show such a dip.
However, the area sampled in each cluster includes a tiny number of
galaxies, so that errors are large and the visibility of a possible dip (if
present) is arguable. The cumulative LF of three clusters at $z\sim0.3$
(Barger et al. 1996)  does not seem to show a dip, but it barely reaches the
dip magnitude.  The only truly comparable LF has been presented by De
Propris et al. (1997), and it is reproduced here in Figure 4 (open squares),
together with the best Schechter fit LF (dotted line) to our data. The
fitting machinery adopted here is discussed in the next subsection. De
Propris et al. (1997) have used Kron magnitudes for faint galaxies and
aperture magnitudes (within a 62 arcsec diameter) for large galaxies (De
Propris 1999, private communication). The two LFs are in remarkable good
agreement ($\chi^2_{\nu}<1$) on the common range ($M_H<-20$ mag), with the
exception of the dip bin, wich is present in our data and absent in De
Propris et al. (1997) data. It is worth to note that the position of the 
dip is well within the spectroscopic sample of De Propris and collaborators,
and thus it could be hardly missed.
The agreement would be even better at $M_H<-24$ if
the De Propris et al. (1997) bright magnitudes were of Kron type, since Kron
magnitudes integrate the galaxy flux inside a smaller area than those
sampled by the 62 arcsec aperture they used.

Since De Propris et al. (1997) studied an almost complementary area of the
Coma cluster with respect our one, and the dip is present in our LF and
absent in theirs, it is possible that the amplitude of the dip depends on
the location in the cluster, as it seems to be the case in the optical
(Sekiguchi 1998). Since the $H$ band luminosity traces the galaxy mass, as
stressed in the Introduction, the possible dependence of the dip amplitude
on the cluster location points out a dependence of the mass function on the
surveyed region, possibly due to a joint effect of morphological dependence
of the LF and variation of the morphological composition over the Coma
cluster. A similar trend is seen in the optical (Andreon 1998). 
Such differences in the LF as a function of the location in the cluster
could be related to subcluster structure. Several evidences for
cluster-cluster merger are present in Coma. Two main peaks appear in the 
X-ray flux density (White et al. 1993), in the projected distribution of 
galaxies (Fitchett \& Webster 1987, Mellier et al. 1988) and in the radio 
source counts (Kim et al. 1994): a clump centered on NGC4874 and NGC 4889,
and a secondary peak around NGC 4839, about 40' SW from the previous one. 
The field 
surveyed here is centered $\sim 15'$ NE from the main structure, at
the opposite side with respect to the cluster center. Colless et al. (1993)
have shown the complex dynamics and multiple substructure of the Coma cluster 
using a large redshift catalog. According to them, the NGC 4839 group 
is actually falling into the main cluster, there are two subclusters in the
central region (associated with the two dominant galaxies), and  
late type galaxies are falling into the main cluster (which is dominated by 
early type galaxies). These processes might be able to locally modify the
LF as observed. 


\subsection{Fitting the LF}

Let us consider now the overall shape of the LF.  Usually, a $\chi^2$ method
is used to fit the LF of clusters by a Schechter (1976) function:

$ f(m)=\phi^* \ 10^{0.4 (\alpha+1)(m^*-m)} \ exp(-10^{0.4(m^*-m)})$

The $\chi^2$ method is not the optimal one for fitting a function to a small
number of bins, and it is even less suitable when bins are poorly populated.
Furthermore, a $\chi^2$ requires to bin the data with an arbitrary bin size.
Although the $\chi^2$ method is not optimal, we are forced to use it, since
we do not know any other fitting method that could take into account, even
roughly, background fluctuations together with Poissonian ones without
binning the data. More elegant methods implemented so far, such as
maximum--likelyhood fitting, do not take into account neither Poissonian
fluctuations of the background counts, nor the field to field variance of
the background, and therefore they systematically underestimate the true
errors.

In order to take into account the amplitude of the bin in the fitting
process (a technical detail seldom considered), we fit the data with a
Schechter function convolved with the bin width (although in practice this
detail makes almost no difference on the results). An additional problem
arises: given the existence of a real dip in the Coma LF, the fit of the
whole LF with any Schechter function is necessarily poor (and in fact we
found a minimum $\chi^2$ of 14 for 4 degrees of freedom). We are therefore
left with two options: flag the dip point, or use a more complex function.
Disposing of a very small number of points and lacking any physically
motivated more elaborate function to be fitted, we simply flag the outlier
bin.

In that case, and taking into account the finite amplitude of the bin, we
found $M^*_H=-24.6$ mag and $\alpha=-1.3$, but with large confidence
intervals (as shown in Figure 3).  Note that our magnitude limit, $H=17$
mag, is roughly equivalent to $M_B\sim-15$ mag at the Coma distance for an
early--type galaxy ($B-H\sim4$ mag), which is well in the dwarf regime. $M^*$
agrees well with the values expected from the optical photometry and usual
colors for early--type galaxies ($M_B=-20.5$ mag and $B-H\sim4$ mag). The slope
is steeper than the typical value in optical bands (L\'opez--Cruz et al.
1997;  Garilli, Maccagni \& Andreon 1999), but nevertheless it is quite
similar to that found for a few well studied clusters in optical
bands (Smith, Driver \& Phillipps 1997; Schechter 1976).


\subsection{Comparison to previous studies of Coma and to the field LF}

Mobasher \& Trentham (1998) studied a very small portion of the Coma cluster
and were able to build a catalog 1.5 magnitudes deeper than ours. However,
their studied field is too small to make the background variance small
relative to the signal (the Coma LF), so that the resulting LF is completely
unconstrained, as admitted by the authors. They computed also another LF,
by performing a crude color selection, i.e.  assuming that Coma cluster
galaxies lay, in a color--color plane, in a region different from that
occupied by the fore and background galaxies. In that case, a LF with
errorbars of reasonable size was derived, but under an hypothesis that
should be demonstrated to be true. In Figure 5, this LF is plotted
overlapped to our $H$ LF, after having matched the two LFs in the common
bins. Their $K$ magnitudes have been changed to $H$ assuming $H-K=0.24$ mag,
the typical value expected for the Coma galaxies. Our errorbars are smaller
in the common bins, even using the same binning for the two LFs.  Mobasher
\& Trentham (1998) points stay relatively near the extrapolation of the best
Schechter fit to our data, suggesting that the LF could keep its
$\alpha\sim-1.3$ slope even at these very faint magnitudes (roughly
equivalent to $M_B\sim-13.5$ mag). The same points stay near the Secker \&
Harris (1996) $R$ band Coma LF shifted in the $H$ band, as plotted in Figure
2.

Figure 5 also compares the Coma cluster LF to the local field LFs, as computed by
Gardner et al. (1997, dashed line) and Szokoly et al. (1998, solid line).  
The two field surveys differ in many respects. The former is based on a
sample about 5 times larger than the latter, and it is computed from a
near--infrared selected sample. Instead, the latter is optically selected
and no corrections have been applied for the optical selection. The slope of
the field LF computed by Gardner et al. (1997) and by Szokoly et al. (1998)
differ largely, with $\alpha\sim-0.9$ and $\alpha\sim-1.3$, respectively.
However, the 68 \% confidence contours of the two LFs cross each other
(figure not shown), implying that the two LFs are compatible to
$\sim1\sigma$, as also claimed by Szokoly et al. (1998). The two field LFs
could have different slopes but they still remain compatible because they
barely reach $M_H=-21.5$ mag\footnote{Both LsF are originally computed in
the $K$ band, and have been transformed in the $H$ band assuming the same
mean rest--frame color for all galaxies ($H-K=0.2$ mag). This is a
reasonable value for the brightest population, which is dominated by
galaxies at $z \ltapprox 0.1$}, and therefore they sample only the
exponentially declining part of the LF.  Therefore, the slope of the field
LF is constrained by the faintest bin (see Figure 5) which, as in all field
surveys, is quite uncertain because measured on a very small volume.

Our own data for the Coma $H$ LF are three mag deeper than the local field
ones, and the overall shape, as parametrized by the Schecther parameters,
agrees with the field ones: the Coma LF has $\alpha$ and $M^*$
indistinguishable from the Szokoly et al. (1998) LF and a $M^*$ very similar
to the Gardner et al. (1997) one.  Its slope, $\alpha=-1.3$, is steeper than
the Gardner et al. (1997) slope, but by less than $\sim1\sigma$ difference,
due to the large confidence level intervals of the two LFs. Also, the
present Coma $H$ band LF agrees with the field $LF$ computed by Cowie et al.
(1996) in a few redshift ranges, up to $z\sim1$.

On one side, the overall shape of the LF is similar both in the field and in
the Coma cluster. On the other side, no dip is present in near--infrared
field LFs, whereas instead in our Coma LF it is quite evident. This fact,
and the absence of the dip in the Coma region studied by De Propris et al.
(1998) seems to suggest that the dip amplitude could be related to the
morphological mix of the studied environment. The alternative possibility
requires that environmental effects change the $H$ band luminosity
preferentially at a given mass (corresponding to $H\sim-22$ mag), without
altering too much the mass distribution for more massive galaxies.
Otherwise, the Coma and the field LFs should have different bright tails.


\section{Discussion \& Conclusions}

We have presented the near--infrared LF of a nearby cluster of galaxies,
Coma, down to faint magnitudes ($M_H=-18.5$ mag, i.e. $M_H^*+6$ mag
corresponding roughly to $M_B\sim-14.5$ mag). This has been provided to be
possible due to the relatively deepness of the present images {\it and} to 
the small background variance associated with the large surveyed area.  The computed
LF is the deepest ever measured in the near--infrared, on any type of
environment.

The shape of the Coma LF in the region studied seems not to depend on 
wavelength, at least in $b$ and $H$ bands, and with less confidence, in $R$. 
The similarity of the LF implies that in the
central region of Coma there is basically no new population of galaxies
which disappears or becomes too faint to be observed in the optical bands
(because of the presence of dust, for instance), down to the magnitudes of
dwarfs. Furthermore, if the $H$ band LF traces the galaxy mass function,
also the blue LF traces the mass in this case. This is in apparent
contradiction with the results by Gavazzi, Pierini \& Boselli (1996), who
found that for spiral galaxies the M/L is approximatively constant in the
near--infrared but not in the optical filters. Since our finding is based on
just one sample in one particular environment, although selected with well
understood selection criteria (volume--complete), it has to be verified on
other samples of nearby galaxies, possibly spiral-rich clusters or groups,
before any dangerous generalization.

The bright part of the Coma $H$ band LF, i.e. the brightest three
magnitudes, agrees with the expectations based on optical LFs and usual
colors for galaxies, and with what is observed in shallower near--infrared
surveys of clusters of galaxies and also on the field. This confirms that
the shape of the tip of the mass function seems environment--independent and
therefore environmental effects have a minor impact on the luminosity of
bright galaxies ($M<M^*+2$), and possibly on their masses. Coma and the field
population differ by a factor of 100 in galaxy density. The extension of this
sentence to faint ($M>M^*+4$) galaxies still await a determination of the
field LF in the dwarf regime.

The Coma near--infrared LF presents a real dip at a luminosity corresponding
to that observed in the optical LF. This is the first detection of such a
feature in the near--infrared. The existence of a dip in the Coma LF in the
$H$ band implies the presence of a dip also in the galaxy mass function.  
To our knowledge, there is presently no simulation of cluster formation
which is able to produce such a feature in the galaxy mass or luminosity 
function. This
feature, being distinctive, will set a strong constraint for the future
simulations.
 
Kauffmann \& Charlot (1998) have shown that the apparent passive evolution
and the slope of the color--magnitude relation can be accomodated within a
hierarchical model, even if the galaxies themselves grow by mergers until
late times. One of the important remaining issues is the comparison between
the predicted and the observed LF, in particular, the distribution of
galaxies as a function of their morphological type, at lest for early--type
galaxies. Probably the main limitation till now has been the lack of
suitable observational data to compare with model expectations. Our
near--infrared catalog, published in Paper I, joint to the morphological
types for the Coma galaxies, available from Andreon et al. (1996, 1997),
fill this observational gap.

The overall slope of the Coma LF is intermediate ($\alpha\sim-1.3$).  The
slope is measured down to the dwarfs regime: we reach $M_B\sim-14.5$ using
own our data alone and even fainter magnitudes ($M_H\sim-17$, roughly
equivalent to $M_B\sim-13$) when including Mobasher \& Trentham (1998) data
and under their assumptions.  When comparing Coma and the field LFs in the
near--infrared, we have to take into account that both LFs have been derived
with completely different data and methods, because of the different 
selection criteria
for the two samples: the field LF is computed on a flux--limited sample,
whereas the cluster LF is computed on a volume--limited sample. In
particular, the field LF suffers from a 10 \% redshift incompleteness (or is
based on an optical selection, as for the Szokoly et al. (1998) LF), and a
poor sampling of faint luminosities, because of the small volume explored at
that luminosities. Nevertheless, and even if the environments sampled are
quite different, the bright tail of the Coma and the field LFs are in close
agreement. In our opinion, this exclude the possibility of large systematic
errors in the derivation of field LFs, and therefore indirectly confirms the
disagreement between the observed near-infrared LF and that expected on
theoretical grounds in the present simulations of a hyerarchical Univers
(Kauffmann, Colberg, Diaferio \& White, 1999). This also suggests that more
detailed models are needed to reproduce the observed properties of galaxies,
such as the LF. In particular, as mentioned before, the existence of a dip in
the present LF and the large range on which this LF is computed (7 mag)
provides a strong constraint to future simulations. It is worth to note that
theoretical predictions on the behaviour of high order statistics, such as
the color distribution or the galaxy evolution, use a particular realization
of the LF as a ``weight". Thus, increasing the accuracy on the determination
of the LF will certainly contribute to the improvement of the theoretical
knowledge on galaxy formation and evolution.

\acknowledgements
Maurizio Paolillo is warmly thanked for sharing with us
the fitting code developed together.  
Discussions with Lucio Chiappetti help to clarify several
fortran subtilities. We are grateful to our collaborators 
in the near--infrared
Coma cluster projet, E. Davoust \& P. Poulain.

\begin{figure}
\epsfysize=8cm
\centerline{\epsfbox[40 195 460 590]{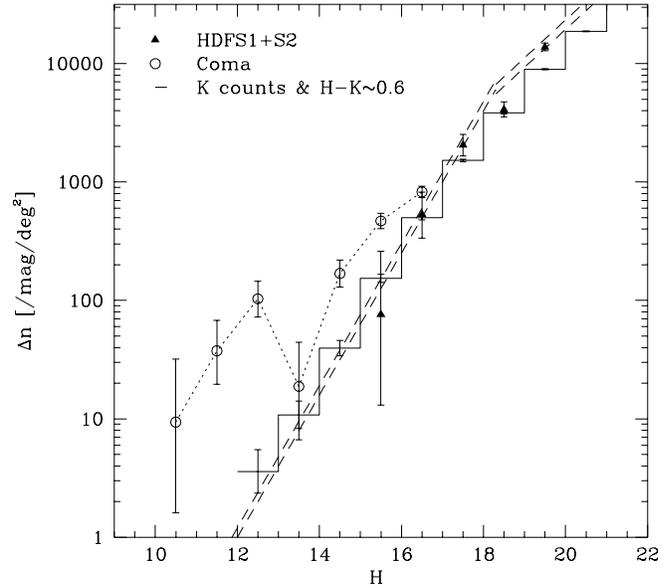}}
\caption[h]{
Galaxy counts as a function of the apparent H magnitude.  Open dots are the
counts in the Coma cluster direction, solid triangles are counts in the HDF
South 1 \& 2 directions. The center of the strip marks the average $K$
counts converted in $H$ assuming $H-K\sim0.6$. The strip width correspond to
a background variance of $\pm15\%$, the typical value for the area surveyed
in Coma. The solid line histogram gives
for comparison the expected counts from our model. See text for details.
Errorbars are computed according to Geherls (1986). Bins are 1
magnitude wide. The abscissa is given by Kron magnitudes for bright galaxies
and aperture magnitudes for faint galaxies.}
\end{figure}

\begin{figure}
\epsfysize=8cm
\centerline{\epsfbox[55 190 460 600]{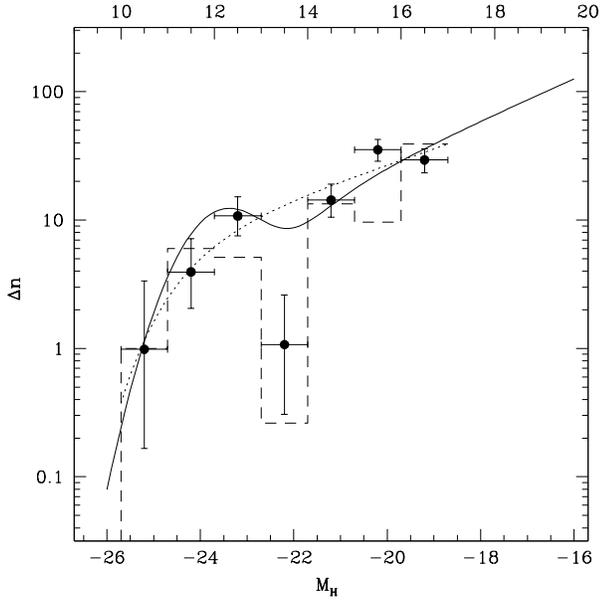}}
\caption[h]{
Coma LF in the $H$ band, as computed from the present data alone (closed
dots).  The solid line is the $R$ Coma LF, shifted by a color $R-H=2.3$
mag. The dashed histogram is the $b$ Coma LF, shifted by a pseudo-color
$b-H=3.5$ mag. The dotted line is the best fit by a Schechter
function, once the dip point is flagged.
Errorbars in the ordinate axis are computed according to
Geherls (1986) and include also the field to field variance of the
background. Errorbars in the abscissa show the bin width.  Errorbars for the
histogram are similar to those of points. The upper abscissa scale shows the
apparent $H$ magnitude, and the lower one gives the corresponding absolute
$H$ magnitude. $\Delta n$ is the 
number of Coma galaxies in the studied field.} 
\end{figure}

\begin{figure}
\epsfysize=8cm
\centerline{\epsfbox[70 190 485 590]{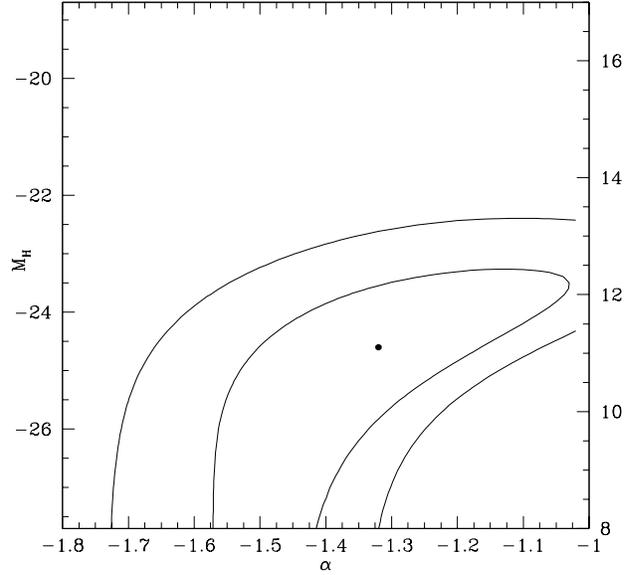}}
\caption[h]{
68 \% and 95 \% confidence contours for the fit of the LF by a
Schechter function. The units of the left and right ordinates
are absolute and apparent $H$ magnitudes, respectively.}
\end{figure}

\begin{figure}
\epsfysize=8cm
\centerline{\epsfbox[60 190 470 600]{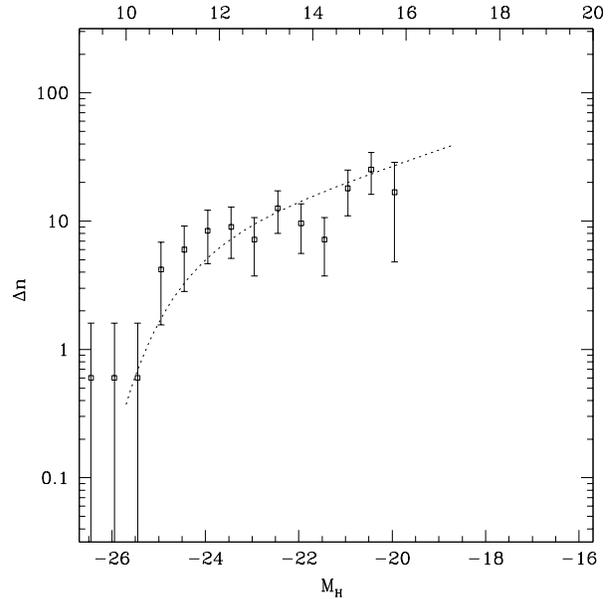}}
\caption[h]{
De Propris et al. (1998) LF (open squares) compared to the best Schechter fit 
to the present data (dotted line).
Errorbars and scales are as described in Figure 2.
}
\end{figure}

\begin{figure}
\epsfysize=8cm
\centerline{\epsfbox[55 190 470 600]{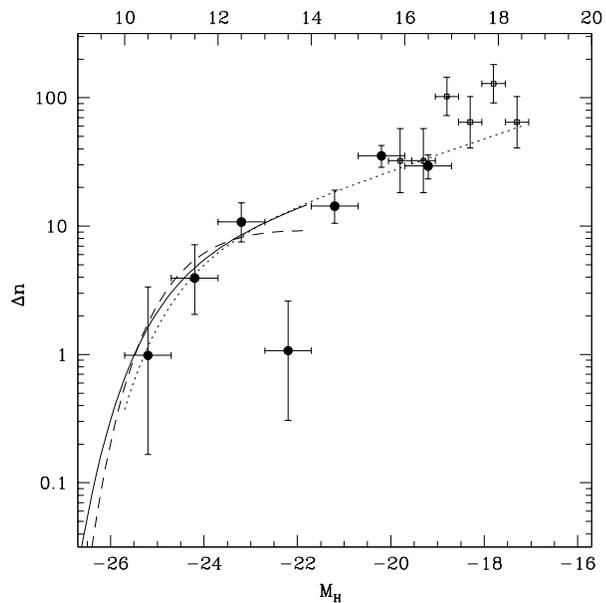}}
\caption[h]{
Various determinations of the near--infrared LF. Our own data (solid dots) and
Mobasher \& Trentham (1998) data (open squares) are shown, after normalization 
of the LF
in common bins. For details on the derivation of the 
Trentham \& Mobasher (1998) data points, see the text.
The dotted curve is the best fit of our data, extrapolated to fainter magnitudes.
Local field LFs are also shown:
Gardner et al. (1997) (dashed line) and  Szokoly et al. (1998) (solid line).
The field LF has been vertically shifted to reproduce 
the Coma LF in the three brightest bins.
Errorbars and scales are as described in Figure 2.
}
\end{figure}
 

\begin{thebibliography}{}

\bibitem[]{}
Andreon S., 1996, A\&A 314, 763

\bibitem[]{}
Andreon S. 1998, A\&A  336, 98

\bibitem[Andreon et al. 1996]{1996A&AS..116..429A} 
Andreon S., Davoust E., Michard R., Nieto J. -L. \& Poulain P., 1996, A\&AS 116, 429 

\bibitem[Andreon Davoust \& Poulain 1997]{1997A&AS..126...67A} 
Andreon S., Davoust E. \& Poulain P., 1997, A\&AS 126, 67 

\bibitem[]{}
Andreon S., Pell\'o R., Davoust E., Dom\'\i nguez R., Poulain P. 1999, A\&AS, in press

\bibitem[]{}
Balogh M. L., Schade D., Morris S. L., Yee H. K. C., Carlberg R. G. \& 
Ellingson E.  1998, ApJ 504, L75 

\bibitem[Barger et al. 1996]{1996MNRAS.279....1B} 
Barger A. J., Aragon-Salamanca A., Ellis R. S., Couch W. J., Smail I. \& Sharples R. 
M., 1996, MNRAS 279, 1

\bibitem[]{}
Bertin E. \& Arnouts S. 1996, A\&AS 117, 393 

\bibitem[]{}
Bershady M. A., Lowenthal J. D. \& Koo D. C. 1998, ApJ 505, 50 

\bibitem[]{}
Biviano A., Durret F., 
Gerbal D., Le Fevre O., Lobo C., Mazure A. \& Slezak E. 1995, A\&A 297, 610 

\bibitem[]{}
Bothun G. D. \& Dressler A., 1986, ApJ 301, 57 


\bibitem[]{}
Bruzual \& Charlot 1993, ApJ 405, 538



\bibitem[]{}
Carlberg R. G., Yee H. K. C., Ellingson E., Abraham R., Gravel P., 
Morris S. \& Pritchet C. J. 1996, ApJ 462, 32 

\bibitem[]{}
Colless, M., \& Dunn, A.~M.~1996, ApJ, 458, 435

\bibitem[]{}
Cowie L. L., Songaila A., Hu E. M. \& Cohen J. G., 1996, AJ 112, 839

\bibitem[]{}
Da Costa L., Nonino M., Rengelink R., Zaggia S., Benoist C. et al.,
1999, A\&A, submitted

\bibitem[]{}
De Propris R. , Eisenhardt P. R., Stanford S. A.  \& Dickinson M.  1998, ApJ 503, L45 

\bibitem[]{}
De Propris R. , Stanford S. A., Eisenhardt P. R.  \& Dickinson M.  1998, AJ in press


\bibitem[]{}
Ferguson H. C. 1989, AJ 98, 367 


\bibitem[]{}
Fitchett M., \& Webster R., 1987, ApJ 317, 653

\bibitem[]{}
Gardner J. P., Cowie L. L. \& Wainscoat R. J. 1993, ApJ 415, L9 

\bibitem[]{}
Gardner J. P., Sharples R. M., Frenk C. S. \& Carrasco B. E. 1997, 
ApJ 480, L99 

\bibitem[]{}
Gardner J. P., Sharples R. M., Carrasco B. E. \& Frenk C. S. 1996, 
MNRAS 282, L1 

\bibitem[]{}
Garilli B., Maccagni D., Andreon S., 1999, A\&A, 342, 408

\bibitem[]{}
Gavazzi G.,  Pierini D. \& Boselli A. 1996, A\&A 312, 397 

\bibitem[]{}
Gehrels N., 1986, ApJ 303, 336

\bibitem[]{}
Godwin J. G. \& Peach J. V. 1977, MNRAS, 181, 323 

\bibitem[]{}
Godwin J. G., Metcalfe N. \& Peach J. V., 1983, MNRAS 202, 113 

\bibitem[Huang et al. 1997]{1997ApJ...476...12H} 
Huang J.-S., Cowie L. L., Gardner J. P., Hu E. M., 
Songaila A. \& Wainscoat R. J., 1997, ApJ 476, 12 

\bibitem[]{}
Impey C. , Bothun G.  \& Malin D.,  1988, ApJ 330, 634 

\bibitem[]{} 
Kauffmann G. , Colberg J. M., Diaferio A.  \& White S. D. M. 1999, 
MNRAS 303, 188 

\bibitem[]{} 
Kauffmann, G.  \&  Charlot, S.  1998, MNRAS 294, 705 

\bibitem[]{} 
Kim, K.~-T., Kronberg, P.~P., Dewdney, P.~E., \& Landecker, T.~L.~1994,
A\&AS, 105, 385

\bibitem[]{}
Kron, R. G. 1980, ApJS 43, 305 

\bibitem[]{}
L\'opez-Cruz O. , Yee 
H. K. C., Brown J. P., Jones C.  \& Forman W.  1997, ApJ, 475, L97 

\bibitem[]{}
Mellier Y., Mathez G., Mazure A., Chauvineau B. \& Proust D. 1988,
A\&A 199, 67

\bibitem[]{}
Mobasher N. \& Trentham B. 1998, MNRAS 293, 315

\bibitem[]{}
Pozzetti L., Bruzual, G., Zamorani G. 1996, MNRAS 281, 953

\bibitem[]{}
Pozzetti L., Madau P., Zamorani G., Ferguson, H., Bruzual, G. 1998,
MNRAS 298, 1133

\bibitem[]{}
Recillas-Cruz E., Carrasco L., Serrano A. et al., 1990, A\&A 229, 64

\bibitem[]{}
Rocca-Volmerange, B., Guiderdoni, B., 1990, MNRAS, 247, 166

\bibitem[]{}
Schechter P., 1976, ApJ 203, 297

\bibitem[]{}
Secker J.  \& Harris  W. E. 1996, ApJ 469, 623 

\bibitem[]{}
Secker J., Harris W. E. \& Plummer J. D., 1997, PASP 109, 1377 

\bibitem[Sekiguchi 1998]{1998ucb..proc...87S} Sekiguchi, M. 1998, 
in Untangling Coma Berenices: A New Vision of an Old Cluster, 
eds. A. Mazure et al., Word Scientific Publishing Co Pte Ltd, p. 87 

\bibitem[]{}
Simard L., Koo D., Faber S., Sarajedini V., Vogt N., Phillips A., Gebhardt
K., et al., 1999, ApJ, in press (astro-ph/9902147)

\bibitem[]{} 
Smith R. M.,  Driver S. P. \& Phillipps S.  1997, MNRAS 287, 415 

\bibitem[]{}
Stanford S. A., Eisenhardt P. R. M. \& Dickinson M.,  1995, ApJ 450, 512

\bibitem[]{}
Szokoly G. P., Subbarao M. U., Connolly A. J., \& Mobasher B.,  1998, ApJ 492, 452 

\bibitem[]{}
Thompson L. A. \& Gregory S. A. 1993, AJ 106, 2197 

\bibitem[]{}
Trentham B. \& Mobasher N. 1998, MNRAS 299, 488

\bibitem[]{}
White S. D. M., Briel U.G. \& Henry J.P., 1993, MNRAS 261, L8

\bibitem[]{}
Williams R.E., Blacker B., Dickinson, M., et al., 1996, AJ 112, 1335

\end{thebibliography}
\end{document}